\documentclass[fleqn,usenatbib,useAMS]{mnras}

\usepackage{graphicx}	
\usepackage{amsmath}	
\usepackage{amssymb}	
\usepackage{multicol}        
\usepackage{bm}		
\usepackage{pdflscape}	
\usepackage{arydshln}

\newcommand{\kms}{\,km\,s$^{-1}$} 

\usepackage[T1]{fontenc}
\usepackage{ae,aecompl}
\defcitealias{Quirk2019}{Q19}

\usepackage{newtxtext,newtxmath}

\title[Asymmetric Drift in the IllustrisTNG Simulation]{Asymmetric Drift of Andromeda Analogs in the IllustrisTNG Simulation}
\author[A. C. N. Quirk]{Amanda C. N. Quirk$^{1}$\thanks{Contact e-mail: \href{mailto:acquirk@ucsc.edu}{acquirk@ucsc.edu}}%
, Ekta Patel$^{2,3}$
\\
$^{1}$UCO/Lick Observatory, University of California at Santa Cruz, 1156 High Street, Santa Cruz, CA 95064, USA \\
$^{2}$University of California, Berkeley, 501 Campbell Hall, Berkeley, CA, 94720 \\
$^{3}$Miller Institute for Basic Research in Science, 468 Donner Lab, Berkeley, CA 94720}

\date{Last updated 2019 December 20}

\pubyear{2020}

\begin{document}
\label{firstpage}
\pagerange{\pageref{firstpage}--\pageref{lastpage}}
\maketitle

\begin{abstract}
We analyze the kinematics as a function of stellar age for Andromeda (M31) mass analogs from the IllustrisTNG cosmological simulation. We divide the star particles into four age groups: $< 1$~Gyr, $1-5$~Gyr, $5-10$~Gyr, and $>10$~Gyr, and compare the kinematics of these groups to that of the neutral gas cells. We calculate rotation curves for the stellar and gaseous components of each analog from 2~kpc to 20~kpc from the center of mass. We find that the lag, or asymmetric drift (AD), between the gas rotation curve and the stellar rotation curve on average increases with stellar age. This finding is consistent with observational measurements of AD in the disk of the Andromeda galaxy. When the M31 analogs are separated into groups based on merger history, we find that there is a difference in the AD of the analogs that have had a 4:1 merger the last 4~Gyr, 8~Gyr, or 12~Gyr compared to analogs that have not experienced a 4:1 merger in the same time frame. 
The subset of analogs that have had a 4:1 merger within the last 4~Gyr are also similar to AD measurements of stars in M31's disk, providing evidence that M31 may in fact have recently merged with a galaxy nearly 1/4 of its mass. Further work using high resolution zoom-in simulations is required to explore the contribution of internal heating to AD.
\end{abstract}




\section{Introduction}
\par The kinematics of stars are influenced by dynamical heating events that occur within (i.e., Giant Molecular Clouds (GMCs), spiral arms, bars) or to (i.e., mergers) the galaxies they reside in. Such events are common, are needed to explain the present-day properties of galaxies, and could explain what makes some galaxies unique \citep[e.g.][]{Seth, Walker1996}. Because of their similar masses ($M_{\rm vir} \approx\ 10^{12}\ M_{\odot}$), the Milky Way (MW) and the Andromeda (M31) galaxies are considered to be analogs of each other \citep[e.g.][]{Patel2017b}. However, despite this similarity, the disk kinematics of the MW and M31 differ drastically. The MW has a population of thin disk stars, while almost all of the disk stars in M31 exist in a thick or kicked up disk \citep{Dorman2013}. These stars are much hotter than classical thin disk stars \citep{Dorman2013}. Additionally, the disk of M31 has a higher velocity dispersion than our own solar neighborhood \citep{Holmberg2009, Sysoliatina2018, Dorman2015, Budanova2017} and a steeper age-velocity dispersion gradient than the MW \citep{Dorman2015, Bhattacharya_2019}.  
While it is possible that the MW and M31 formed under different conditions that led to their distinct dynamics, mergers may also explain the higher velocity dispersion observed in M31 \citep{Leaman2017}. As searching for evidence of dynamical heating via internal and external sources is more easily accessible than disentangling different galactic birth conditions, in this study, we aim to understand the impact of external sources of dynamical heating, specifically that caused by mergers, on the observed kinematics of stellar populations in M31.

\par Mergers lead to the growth of galaxies via the accretion of gaseous and stellar material to the inner and outer regions of galaxies \citep{Abadi2003}; thus they affect not only galactic disks but also the halos of galaxies by perturbing the orbits of stars residing at radial distances of a few to a few hundred kiloparsecs from the center of a galaxy \citep[e.g.][]{Quinn1986}. These dynamically heated stars can lead to a kicked-up or puffed up disk, like the one seen in M31 \citep[e.g.][]{Purcell2010, Dorman2013}. Additionally, simulations show that mergers can heat or accrete stars: these heated or accreted stars can have a permanent increase in velocity dispersion of up to 400\% when compared to stars formed in-situ \citep{Walker1996}. Since stars have long dynamical time scales, they retain these differences \citep[e.g.][]{Abadi2003, Purcell2010}. In addition to heating existing stars, mergers can play an important role in setting the birth kinematics of stars. \citet{Brook_2004} illustrate that stars born in epochs of rapid accretion of gas from hierarchical growth exhibit different and much hotter $z=0$ kinematics than stars that were born during periods of quiescence. Thus, mergers play an important role both in setting the birth kinematics of stars in a merger event and also heating the kinematics of existing stars.

\par Because of its relatively close proximity of 785~kpc \citep{Mcconnachie2005}, we can study the entire disk and halo of M31 using individual, resolved stars and search for evidence of mergers. In the last decade, two major surveys of the M31 region have revealed potential relics of mergers throughout M31's disk and  halo.  The wide-field Pan-Andromeda Archaeological Survey \citep[PAndAS;][]{Ibata_2007, Ibata_2013} has surveyed the halo of M31 to a projected distance of 150~kpc, revealing debris in the form of tidal streams and shelves \citep[e.g.][]{Hernquist1988, Hernquist1989, Escala2019, McConnachie_2018}. The most prominent debris feature is the Giant Stellar Stream (GSS) \citep{Ibata2001, Ibata, Ferguson}, which spans nearly 100 kpc from the center of M31. The second survey is the Pan-chromatic Hubble Andromeda Treasury (PHAT) \citep{Dalcanton2012, Williams2014}, which among other discoveries has revealed M31's global star formation history \citep{Lewis2015, Williams2017}, a ring of star formation at 10~kpc \citep{Lewis2015}, and the age-velocity dispersion relation \citep{Dorman2013}. 

\par As more evidence for a fairly constant and violent merger history for M31 have emerged, so has the question about the types of mergers that shaped M31 into the galaxy we see today. Until recently, the favored theories for M31's formation history were an ancient major merger followed by a series of minor mergers \citep[e.g.][]{vdb_2006, Courteau_2011, Kormendy2013}. Simulations of minor mergers in M31's halo have successfully recovered many observed features: the simulations from \citet{Tanaka2009} show that roughly 15 minor mergers are necessary to form the stellar debris seen in M31 today, and \citet{Fardal2008} illustrates that a minor merger can recover the formation and velocity dispersion of the GSS as well. Furthermore, the motions of globular clusters at high galactic radii suggest that M31 has had more than one major period of accretion \citep{Mackey2019}. However, new studies suggest that instead of an ancient and larger merger, a relatively recent ($\sim$2-3 Gyr ago) major merger could explain many of these observed phenomenon while preserving the disk of M31 \citep{DSouza2018}. For example, \cite{Hammer2018} find a major merger model recovers both the substructure resembling the GSS and the age-velocity dispersion gradient in M31's disk \citep{Dorman2015}. They conclude a major merger can explain the burst of star formation approximately 2~Gyr ago \citep{Williams2017}, a ring of star formation at 10~kpc \citep{Lewis2015}, and the mass of the stellar halo \citep{Ibata_2013, Bell2017}. Thus, while it is believed that M31 has had a long and ongoing history of accretion \citep[e.g.][]{Ferguson2016}, the exact merger history is still unknown, and therefore leads to an ongoing debate as to whether or not a single, recent major merger could explain several observed properties in M31's halo and disk better than a series of recent minor mergers. 

\par \citet[][]{Hammer2018} highlight the importance of recovering M31's disk dynamics when testing different merger histories for M31. \emph{In this study, we aim to determine if the relative difference between the stellar dynamics and gas dynamics of the disk of M31 are consistent with a major merger scenario.} To do so, we will analyze the stellar and gas dynamics across a sample of M31 mass analogs selected from a cosmological simulation, accounting for a range of merger histories across a large sample of simulated galaxies, and compare these properties to observed measurements of the lag between the gas rotation velocity and that of the stars, or the asymmetric drift (AD), in the disk of M31 from \citet[][hereafter \citetalias{Quirk2019}]{Quirk2019}. In doing so,  we will explore whether AD changes with merger history and thereby determine how significant of a role external dynamical heating has played in the current kinematics of these analogs.

\par In this study, we are solely limited to exploring the effects of external sources of dynamical heating (i.e. mergers) because the resolution of simulated analogs selected from a cosmological box does not allow us to probe internal sources occurring on sub-kiloparsec scales such as perturbations from spiral waves \citep[e.g.][]{Sellwood2002, Sellwood2013}, galactic bars \citep[e.g.][]{Dehnen1998, Saha_2018}, or GMCs \citep[e.g.][]{Jenkins2019, Ting2019}. As such, we cannot comment on the effects of these drivers on the disk dynamics of massive galaxies. While internal sources of heating affect a galaxy's kinematics, they cannot alone explain the kinematics observed today. External sources of dynamical heating, mergers, are also necessary to explain some Local Group galaxies' velocities \citep{Leaman2017}. 
Furthermore, while stellar feedback can drastically affect the disk of galaxies less massive than the MW or M31 \citep{El_Badry2016}, mergers are more important than stellar feedback for setting the present-day kinematics of massive galaxies. Thus, in this work we focus solely on the effects of external dynamical heating on AD, with the overarching goal of shedding light on the recent merger history of M31.

\par This paper is organized as follows: in Section~\ref{sec:Data}, we discuss the IllustrisTNG simulations and the selection criteria we use to choose our sample of analogs. Section~\ref{sec:Methods} describes the methods used to calculate rotation velocities, construct rotation curves, and calculate AD. In Section~\ref{sec:Results}, we illustrate how AD as a function of stellar age changes with merger history, and in Section~\ref{sec:Discussion} we make comparisons to observational results. In Section~\ref{sec:Summary}, we summarize our findings.

\section{Data} \label{sec:Data}
\subsection{The Illustris Suite}
The IllustrisTNG Project is a suite of N-body and hydrodynamic simulations evolved from redshift $z=127$ to $z=0$ using the moving-mesh \texttt{AREPO} code \citep{marinacci18, naiman18, springel18, pillepich18, nelson18, springel10}. The simulations span a cosmological volume of (110.7 Mpc)$^3$ and are initialized with the following cosmological parameters from \cite{planck15}: $\Omega_m= 0.3089$, $\Omega_{\Lambda}=0.6911$, $\Omega_{b}=0.0486$,  $\sigma_8=0.8159$, $n_s=0.9667$, and $h=0.6774$. Throughout our analysis, we adopt a value of $h=0.704$ \citep[WMAP-9;][]{hinshaw13}.

 The IllustrisTNG Project \citep{marinacci18, naiman18, springel18, pillepich18, nelson18} is a follow-up suite of simulations to the original Illustris Project \citep{nelson15,vogelsberger14}, including an updated galaxy formation model described in \citet{weinberger18} and \citet{pillepich18}. The most significant improvements to the galaxy formation model include the treatment of AGN feedback, an improved parametrization of galactic winds, and the addition of magnetic fields.

In our analysis, we use data from the IllustrisTNG100-1 simulation (hereafter IllustrisTNG), the midsize and mid-resolution simulation in the suite including both dark matter and baryons. IllustrisTNG follows the evolution of 1820$^3$ dark matter particles and 1820$^3$ hydrodynamical cells, achieving a dark matter particle mass resolution of $m_{\rm DM}=7.5 \times 10^6 \, M_{\sun}$ and a baryonic mass resolution of $m_{\rm bary} = $ $1.4 \times 10^6 \, M_{\sun}$. Halos and subhalos are identified using the \texttt{SUBFIND} \citep{springel01, dolag09} halo-finding routine. We use the IllustrisTNG merger trees created with the \texttt{SUBLINK} code \citep{rg15} to track the mass history of subhalos in our analysis. In the following sections, we identify a primary sample of M31 analogs in IllustrisTNG.

\subsection{Selection of M31 Mass Analogs} 
\label{subsec:M31analogs}
M31 analogs are chosen as all central (or primary) subhalos at $z=0$ where the FoF group virial mass is $M_{\rm vir}=1-2.2 \times 10^{12}~M_{\sun}$ \citep[see][]{DSouza2018}. Virial mass is calculated for each \texttt{SUBFIND} FoF group following the methods of \citet{brynorman98}. We also require that these primary subhalos have a stellar mass within the range $5-20\times10^{10}~M_{\sun}$ \citep{tamm12, sick2014}. Finally, we require that the maximum circular velocity for each subhalo is at least 200 $km~s^{-1}$ to ensure that our analogs generally represent the observed HI rotation curve of M31 \citep{Chemin2006, corbelli10}. Using these selection criteria, we find 216 M31 analogs in IllustrisTNG. We eliminate one analog that has no associated gas, bringing the \emph{full sample} to a total of 215 analogs.

\begin{figure*}
\begin{center}
\includegraphics[scale=0.5, trim=3.5cm 5mm 2cm 0mm]{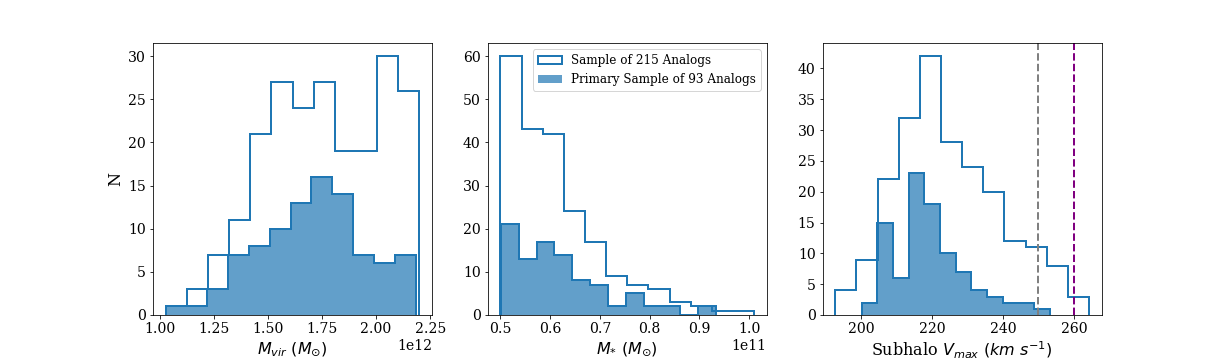}
\caption{Histograms of virial mass (left), stellar mass (middle), and subhalo $\rm v_{max}$ for all M31 mass analogs. In all panels, the blue histogram represents the full sample of 215 analogs, and the blue filled histogram represents our primary sample of 93 analogs that are chosen based on visual inspection of their radially average rotation curves (see Section \ref{subsec:rotcurve}). In the right panel, the gray vertical dashed line denotes the peak circular velocity of M31 as measured in \citet{corbelli10} and the purple dashed line represents the approximate peak circular velocity of both halves of M31's disk in \citet{Chemin2006}. Our primary sample is generally representative of the full sample of M31 analogs in all properties. \label{fig:analog_masses}}
\end{center}
\end{figure*}

The left panel of Figure \ref{fig:analog_masses} shows the distribution of virial masses for the full sample (blue histogram) and for the subset of 93 M31 analogs that will serve as the \emph{primary sample} (blue filled histogram). The primary sample is selected based on visual inspection of each simulated galaxy's rotation curve and provides a checkpoint to eliminate any analogs without sufficient statistics to calculate both stellar and gaseous rotation curves (see Section \ref{subsec:rotcurve}). The middle and right panels show the distribution of stellar mass and maximum circular velocity, respectively, for the full sample of M31 analogs in IllustrisTNG (blue histograms) and the primary sample (blue filled histograms). The gray vertical line in the right panel indicates the maximum circular velocity of M31's HI rotation curve \citep{corbelli10}. All three properties of the primary sample are representative of the broader population of M31 analogs.

\subsection{Stellar Assembly for IllustrisTNG Analogs}
\label{subsec:stellarassembly}

Our main goal is to explore if the merger history of galaxies in our simulated sample show any correlation with the values of AD. As such, the stellar assembly and merger history catalogs for the IllustrisTNG suite are also used to probe the formation histories of the M31 analogs identified in Section \ref{subsec:M31analogs}. The stellar assembly and merger histories of the original Illustris-1 simulated galaxies are discussed in \citet{rg15,rg16a, rg16b}. The same tools were used to create the equivalent catalogs for IllustrisTNG. Below, we describe the properties of interest for this analysis and their definitions as defined by the previously mentioned papers. 

\begin{itemize}
    \item time of last major merger: the lookback time at which the last major merger (stellar mass ratio $<$ 4:1) occurred
    \item total number of major mergers: the total number of completed major mergers (stellar mass ratio $<$ 4:1)
    \item total number of minor mergers: the total number of completed minor mergers (stellar mass ratio $>$ 10:1 and $<$ 4:1)
\end{itemize}

\begin{figure}
\begin{center}
\includegraphics[scale=0.6, trim=0mm 0mm 0mm 0mm]{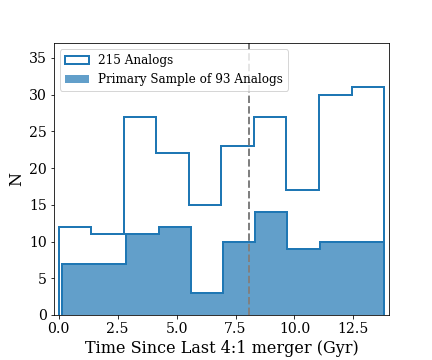}
\caption{Histograms of time since an analog's most recent 4:1 merger. The blue histogram represents the full sample of 215 analogs, and the blue filled histogram represents our primary sample of 93 analogs. The vertical line marks the median time for the primary sample at 8.0 Gyr. Our primary sample is generally representative of the broader distribution of M31 analogs.\label{fig:merger_times}}
\end{center}
\end{figure}

Figure \ref{fig:merger_times} shows the distribution of the time of last major merger for the 215 M31 analogs in our primary sample. The median time of last major merger is 8 Gyr ago as denoted by the vertical dashed gray line, thus many analogs have had relatively quiet recent merger histories. 

\begin{figure}
\begin{center}
\includegraphics[scale=0.6, trim=0mm 0mm 0mm 0mm]{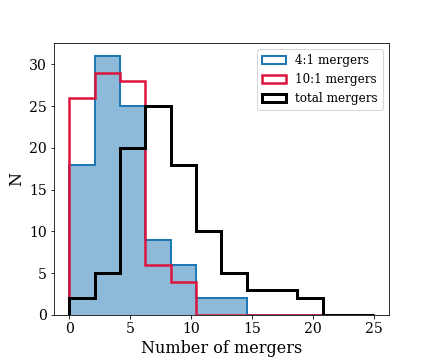}
\caption{Histograms of number of mergers for the analogs in the primary sample. The blue filled histogram shows the distribution of number of 4:1 mergers, the red histogram represents the number of 10:1 mergers, and the black histogram represents the total number of mergers.  \label{fig:analog_num_mergers}}
\end{center}
\end{figure}

Figure \ref{fig:analog_num_mergers} illustrates the total number of mergers (major and minor) for the primary sample in black. The distribution of total 4:1 mergers is indicated by the blue filled histogram and the total number of 10:1 mergers is represented by the red histogram. On average, M31 analogs in our primary sample experience 5-10 mergers in total. In total, the primary sample experiences 441 4:1 mergers, 369 10:1 mergers, summing to 810 mergers overall.

\section{Methods}
\label{sec:Methods}
AD is often defined as the difference between the circular velocity derived from the potential and the stellar rotation velocity \citep{Stromberg}. For the purposes of comparing this work to \citetalias{Quirk2019}, we define AD slightly differently than \citet{Stromberg} and \citet{Walker1996}. Instead of relying on models of the potential to calculate circular velocity, we use the rotation velocity of the gas data to calculate AD, as in \citetalias{Quirk2019}, since this gives a more empirical measurement. 

The rotation velocity of the gas is used as a proxy for circular velocity because gas is collisional and therefore can easily dissipate energy and maintain a low energy orbit \citep{Sellwood11999}. Gas creates radial pressure support, which in low mass disks can cause the tangential velocity of the gas to fall below its circular velocity; however this is not the case for massive disks like M31 \citep{Dalcanton_2010}. Stars, on the other hand, are not collisional and thus can maintain an eccentric orbit if they are gravitationally perturbed. When comparing their circular rotation velocities, stars may lag behind gas because they are following a different orbital path and/or moving more slowly. 

\par In this sense, AD is a measurement of the difference between a star's orbit and a roughly circular orbit, as the gas is only a proxy for circular velocity. Thus, it can also be treated as a long-term effect of dynamical heating. Since the perturbations from dynamical heating can permanently alter the orbital paths of stars \citep[e.g.][]{Leaman2017}, it can change a star's angular momentum such that random motions are increased in a stellar population, and thus the group's average motion is less dominated by orderly rotation \citep{Sellwood2002}. Effects of dynamical heating may be long lasting, and individual stars in galaxies often preserve the effects of heating. \citet{Walker1996} find that minor mergers temporarily increase AD in a galaxy using idealized N-body simulations. 
\par In this work, we will focus on the effect of external dynamical heating, and specifically that caused by 4:1 major mergers for a cosmological sample of M31 mass analogs. Where possible, we will make comparisons to existing AD measurements in M31, as this is the only galaxy for which observational results of AD as a function of stellar age are available \citepalias{Quirk2019}. In our own galaxy, AD has been used to construct rotation curves beyond the solar neighborhood and to correct the local standard of rest (LSR) \citep{Golubov2014, Huang2015} but has not been studied as a function of age or for individual stars.

\subsection{Rotating the Coordinates}
\label{subsec:rotcoord}
To calculate rotation curves and subsequently AD, we first apply a center of mass (COM) shift and rotate the particles/cells in each simulated M31 analog to a face-on reference frame. IllustrisTNG particle/cell positions and velocities are reported relative to the simulation box edges, so we first shift the data to a (0,0,0) Cartesian reference frame by subtracting the COM of the dark matter halo from both the stellar and gaseous data. This COM position comes directly from the IllustrisTNG group catalogs (i.e. the position of each FoF group). We have checked whether using the center of mass of the baryons rather than the dark matter alters the results and find no significant differences.

To rotate the particle/cell data to a face-on reference frame, we calculate the net specific angular momentum vector for all gas cells belonging to a given halo. Then we apply a rotation matrix to the cell data such that the unit vector of the net angular momentum of the gas is mapped to the z-axis in the face-on reference frame. The same procedure is repeated with all star particles belonging to a given halo. Thus, the stars and gas are each rotated with respect to their own net angular momenta. In this new reference frame, the galaxy's disk lies in the xy-plane. Finally, particles with a z scale-height $>$ 10 kpc are eliminated to ensure we only include data for the gas and stars roughly in the disks to construct rotation curves and subsequently AD.

\subsection{Smoothing the Particles/Cells}
\label{subsec:smoothing}
Once the particles/cells are rotated to the face-on reference frame, we divide the stellar particles into four broad age groups: Group 1 with ages $< 1$~Gyr, Group 2 with ages $1-5$~Gyr, Group 3 with ages $5-10$~Gyr, and Group 4 with ages $>10$~Gyr. These age bins differ from those used in the observational analysis of AD in M31 in \citetalias{Quirk2019} in order to take advantage of the full range of stellar particle ages that exists in the IllustrisTNG simulation. The observational constraints on M31 in \citetalias{Quirk2019} are limited to stellar ages of up to $\sim$4 Gyr, as older stars are too faint to observe easily. IllustrisTNG, on the other hand, provides ages spanning approximately a Hubble time. However, there is a lack of young ($<500$~Myr) stars in IllustrisTNG, which prevents us from breaking Group 1 into age bins that better match the observational results in M31. This lack of young stars may be a result of the AGN feedback model in IllustrisTNG \citep{Weinberger2018, Terrazas2020}.

\citetalias{Quirk2019} calculates AD of stars with respect to neutral {\sc Hi} gas. To follow this, we also calculate AD with respect to neutral gas data in this analysis. With the IllustrisTNG simulated galaxies, we implement this by limiting our analysis to only include the gas data where the neutral hydrogen fraction, $n_H $, is above some threshold. Since there is no exact boundary for what gas constitutes as neutral, we initially varied the neutral fraction cutoffs to find the highest fraction that still included enough gas cells for good statistics and to have a full spatial expanse across the inner 20~kpc of each analog. We find that $n_H >$ 0.6 is a reasonable threshold, as on average $21.1\pm11 \%$ of gas particles fall into this range for the primary sample.

\par After the stellar particles are divided into the four age groups and the neutral gas cells are isolated, we locally smooth the line-of-sight velocity (the vertical component or $V_z$ in the face-on reference frame). Instead of smoothing velocities using a grid of set positions, we use the true positions of all particles/cells in a given M31 analog and center a smoothing circle on each particle/cell, as in \cite{Dorman2015} and \citetalias{Quirk2019}.For stellar age Groups 2 and 3, we use a circle of radius 0.75~kpc and a circle of 1.05~kpc for the less populated Groups 1 and 4 and for the neutral gas cells. This radius size mimics the 200~\arcsec-275~\arcsec resolution in {\sc Hi} observations in M31, described in \citetalias{Quirk2019}. The size of the circle was optimized to probe local kinematics and to include enough data points for good statistics. If there are at least 10 particles/cells in a circle, we assign the median of the velocity components to that particle/cell. We use these median components to calculate a rotation velocity and assign it to the position of the particle/cell that the circle is centered on. If there are not 10 members in a particle's/cell's smoothing circle, we do not calculate a rotation velocity at that position. That particle/cell may still be included in the membership of nearby circles and thus would contribute towards the calculation of the median velocities at other positions if those circles pass the membership requirement. We also experimented with adjusting circle membership requirements to balance retaining data with being able to smooth the data. Similar to \citetalias{Quirk2019}, we chose a membership requirement of 10.

To ensure that our smoothing does not obscure any substructures in the rotation curves and subsequent bias to AD, we created rotation curves with and without the smoothing and find no significant differences. We illustrate the smoothing process for an example halo in our primary sample in Section~\ref{subsubsec:smoothing}.

\subsection{Calculation of Rotation Curve}
\label{subsec:rotcurve}

For the rest of our analysis, we use the smoothed kinematics for the primary sample of M31 analogs. We calculate a rotation velocity for every star particle and gas cell that passes the circle-membership requirement in the smoothing process. To calculate the rotation velocity, we find the tangential velocity of the particles/cells by first calculating the projected radial velocity and the total planar velocity, shown in the equations below. We limit our rotation velocity calculations to the 2D xy plane in our face-on rotated frame so the resulting rotation velocities are comparable to de-projected rotation velocities in observations \citepalias{Quirk2019}.  
\begin{equation}
    v_{rad} = \frac{x \cdot v_{x} + y \cdot v_{y}}{\sqrt{x^{2} + y^{2}}}
\end{equation}

\begin{equation}
    v_{tot} = \sqrt{v_{x}^{2} + v_{y}^{2}}
\end{equation}

\begin{equation}
    v_{rot} = v_{tan} = \sqrt{v_{tot}^{2} - v_{rad}^{2}}
\end{equation}

\par The above procedure gives us a rotation velocity for every smoothed star particle and neutral gas cell. To calculate AD  we must be able to directly compare the rotation velocity of the star particles and neutral gas cells. However, the gas cells and star particles do not have the same spatial extent and analogs typically have an unequal number of stellar particles and gas cells. Instead of trying to pair each star particle to a neutral gas cell as in \citetalias{Quirk2019}, we radially bin all of the particles/cells and calculate the median rotation velocity within that bin. Our radial bins sample from 2~kpc to 20~kpc with a spacing of 0.1~kpc. We radially bin in this fashion separately for each stellar age bin and also for the neutral gas cells. As seen in the bottom panel of Figure~\ref{fig:analog_map}, there is no clear azimuthal dependence of the rotation velocity, so we do not obscure any local substructure through this binning. 

\subsubsection{Visual Inspection of Rotation Curves}\label{subsec:visual_inspec}

Since both the gas cells and star particles of these M31 analogs are necessary to calculate AD, we perform an additional visual inspection of the rotation curves of each analog in the full sample to ensure they have enough spatially overlapping stellar particles and gas cells across the inner 20~kpc of the analog. Those that do not have spatial overlap between the gas cells and star particles for at least five radial bins are removed. After this visual inspection, 93/215 M31 analogs from IllustrisTNG remain and will be referred to as the \emph{primary sample}. 
\subsubsection{Demonstrating Smoothing Particle/Cell Data}
\label{subsubsec:smoothing}

\begin{figure}
    \begin{center}
    \includegraphics[width=\columnwidth, scale=7]{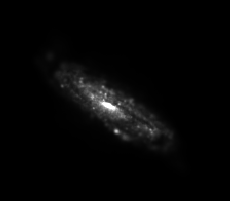}
    \caption{Mock image of analog 449972 as if viewed with the Sloan Digital Sky Survey (SDSS) {\it{g}}  filter.}
    \label{fig:mock}
    \end{center} 
\end{figure}
\begin{figure*}
\begin{center}
\includegraphics[scale=0.45]{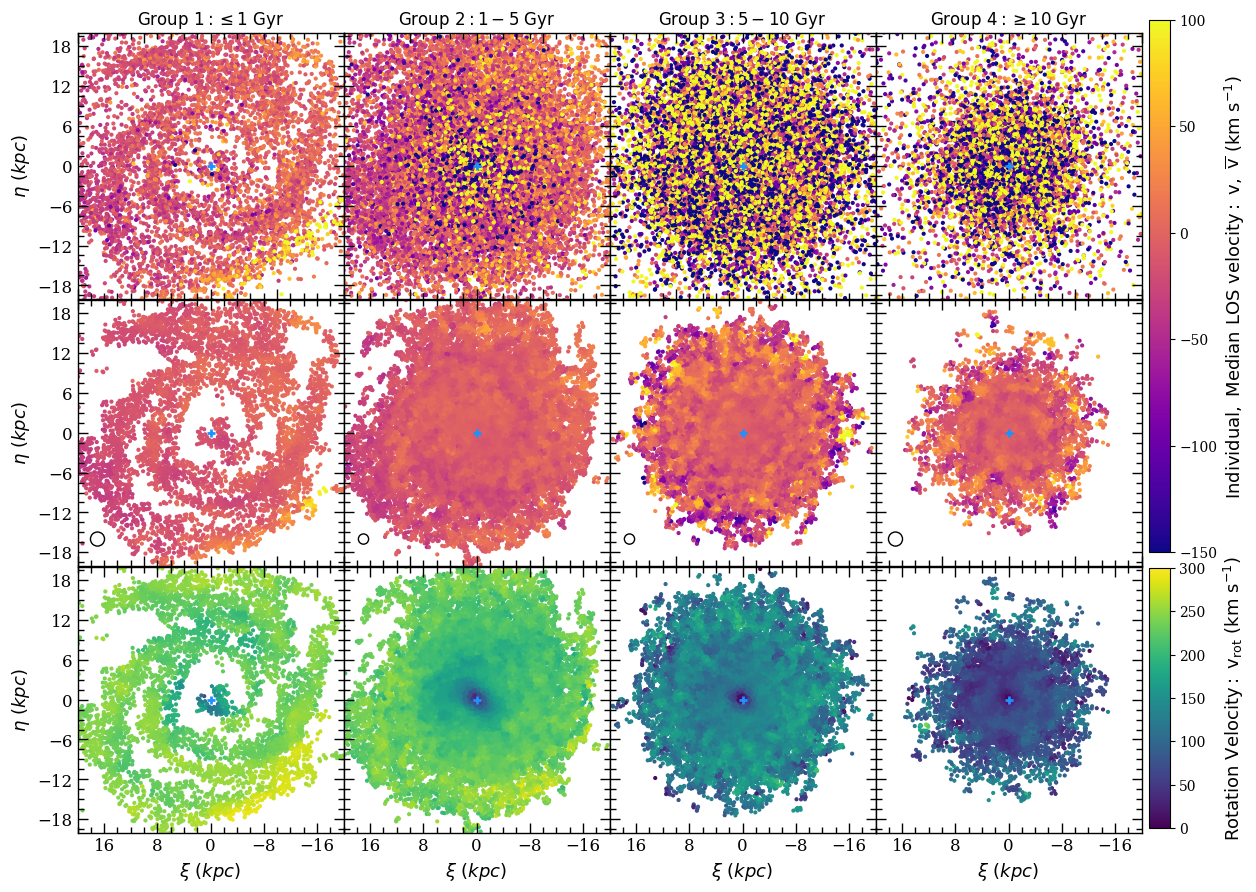}
\caption{Individual line-of-sight velocity (top row), local medians of the line-of-sight velocity (middle row), and rotation velocity (bottom row) as a function of location and age for star particles in analog 449972. From left to right: age $< 1$~Gyr, age $1-5$~Gyr, age $5-10$~Gyr, and age $>10$~Gyr. Smoothing circles are used to calculate the median of the individual line-of-sight velocities. We use a circle of radius 0.75~kpc for Groups 2 and 3 and 1.05~kpc for the less populated Groups 1 and 4. The circles in each panel of the middle row show the respective sizes of the smoothing circles. The blue cross marks the center of the analog. \label{fig:analog_map}}
\end{center}
\end{figure*}

In this section, we choose an analog from the primary sample that best represents M31's rotation curve to demonstrate smoothing the particle/cell data prior to calculating rotation curves. The halo properties of this example analog are listed in Table~\ref{tab:analog}, and the mock image of this simulated galaxy is shown in Fig~\ref{fig:mock} \citep{Torrey2015, Rodriguez-Gomez19}. The results of velocity smoothing (described in Section \ref{subsec:smoothing}) for this analog is illustrated in Figure~\ref{fig:analog_map}, which shows the individual line-of-sight velocity, the post-smoothing median line-of-sight velocity, and the rotation velocity of each star particle for the four age bins as a function of position. For this analog, 16088 of the original 19919 central neutral gas cells remain after the smoothing process, and 77832 of the 87141 central star particles remain. This is typical of M31 analogs in the primary sample. The smoothed kinematic data retains the underlying kinematics needed to calculate AD  without strong loss of  information, so we continue to smooth the data for the entire primary sample.

\begin{table}
\caption{Analog 449972 Halo Properties}
\begin{tabular}{lr}
     \hline
     $M_{\rm vir}$ & $2.01\times 10^{12}~M_{\sun}$ \\[2.5mm]
     $M_{\star}$ & $9.08\times 10^{10}~M_{\sun}$ \\[2.5mm]
     $V_{\rm max}$ & $247~ \rm km~s^{-1}$ \\[2.5mm]
     Time Since Last 4:1 Merger & 4.12~Gyr \\[2.5mm]
    Number of 4:1 Mergers & 4 \\[2.5mm]
     Number of 10:1 Mergers & 5 \\
     \hline
\end{tabular}
\label{tab:analog}
\end{table}

\subsection{Calculation of AD as a function of Stellar Age}
AD, (also denoted by $v_a$), is the difference between the rotation velocity of the gas and that of the stars. 

\begin{equation}
    v_a = v_{\rm rot,gas} - v_{\rm rot,\star}
\end{equation}

\begin{figure*}
\begin{center}
\includegraphics[scale=0.46]{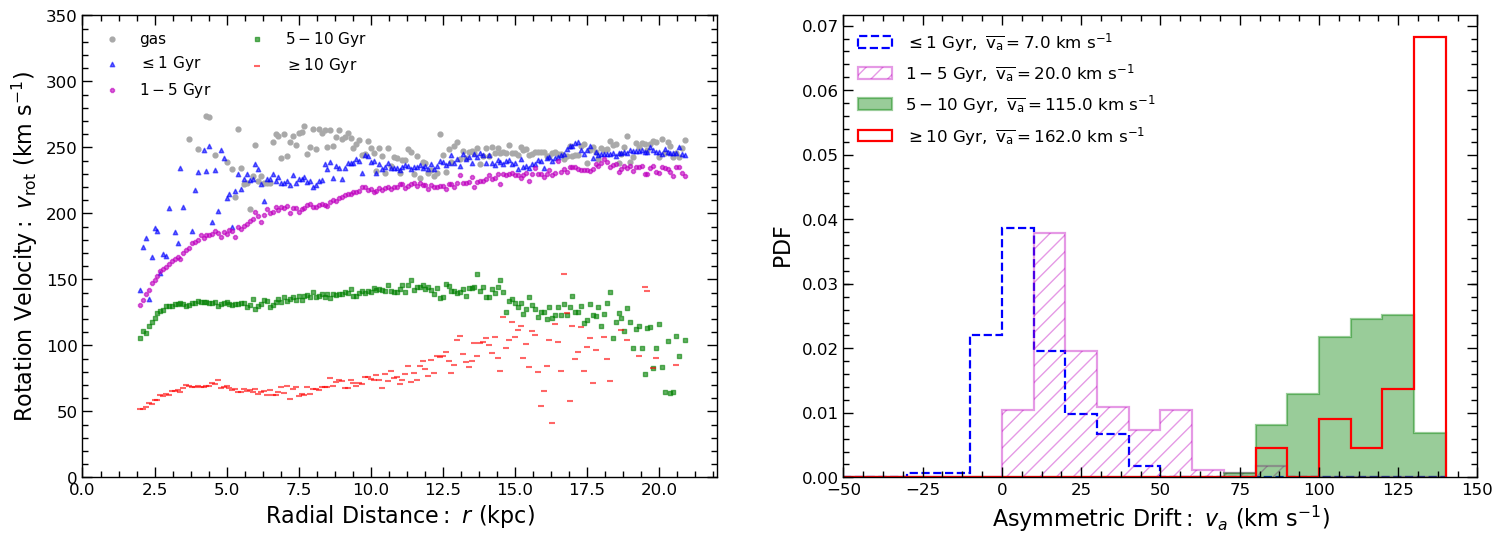}
\caption{Left panel: Radially binned rotation curve for analog 449972. The neutral gas is represented by the grey points, the blue triangles represent stars with ages $< 1$~Gyr, the purple points represent stars between $1-5$~Gyr, the green squares represent stars between $5-10$~Gyr, and the red dashes represent the stars $>10$~Gyr. Right panel: normalized histograms (probability distribution function) of AD values for analog 449972. The blue dashed histogram represents stars that are $< 1$~Gyr, the one with the purple hatch marks represents stars between $1-5$~Gyr, the green solid histogram represents stars between $5-10$~Gyr, and histogram denoted by the red line represents the stars $>10$~Gyr. AD increases with stellar age. \label{fig:analog_ad}}
\end{center}
\end{figure*}

We calculate the AD at each radial bin for the four stellar age groups individually. Thus at each radial bin, we have an AD measurement for Group 1, 2, 3, and 4 with respect to the same neutral gas cell rotation velocities. An example of the results of the AD calculations from the corresponding rotation curve are presented in Figure~\ref{fig:analog_ad}. The left panel shows the rotation curve for this analog: the neutral gas is represented by the grey points, the blue triangles represent stars with ages $< 1$~Gyr, the purple points represent stars between $1-5$~Gyr, the green squares represent stars between $5-10$~Gyr, and the red dashes represent the stars $>10$~Gyr. The right panel shows histograms of the AD for the four stellar age groups for the example analog 449972. The blue histogram represents stars that are  $< 1$~Gyr, the purple hatch marks represents stars between $1-5$~Gyr, the green solid histogram represents stars between $5-10$~Gyr, and the red histogram represents the stars $>10$~Gyr. In all following figures of AD, the histograms will follow the same style and color pattern for each stellar age bin. The trend of AD increasing as a function of stellar age is clearly visible in this example. In the next section, we describe trends with AD for the primary sample of analogs. 

\section{Results}\label{sec:Results} 
\begin{figure*}
\begin{center}
\includegraphics[scale=0.46]{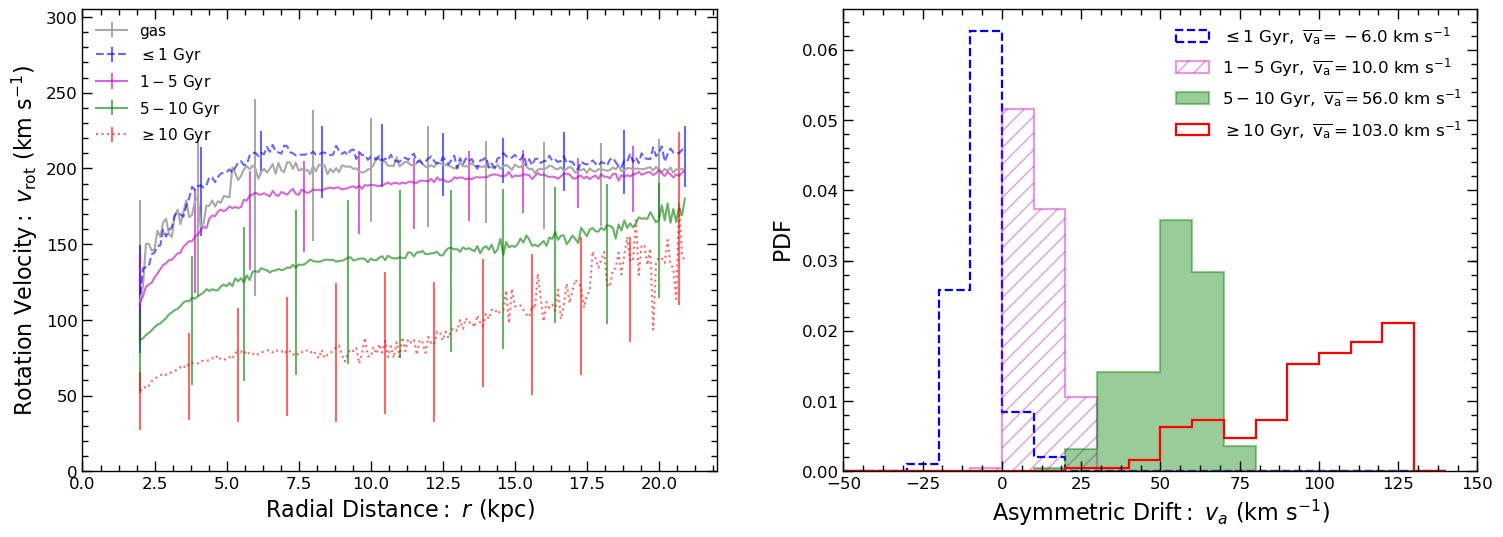}
\caption{Left panel: Rotation curve for the primary sample of analogs. Each point represents the median rotation velocity at a radial bin with a width of 0.1~kpc across all of the analogs in the primary sample. The grey line represents the gas cells, the blue dashed line represent star particles with ages $<1$~Gyr, the purple line represents star particles with ages 1-5~Gyr, the green line represents star particles with ages 5-10~Gyr, and the red dotted line represents star particles with ages $>10$~Gyr. The vertical bars represent the width of the distribution ($16^{\rm th} - 84^{\rm th}$ percentile) of rotation velocities in a given radial bin. Right panel: Normalized histograms (probability distribution function) of AD values for the 93 analogs of the primary sample. For each analog, we find the median AD value of the four age bins; these histograms represent the distribution of those medians across all of the analogs. The colors correspond to the age bins as follows: $< 1$~Gyr (blue dashed histogram), $1-5$~Gyr (purple hatched histogram), $5-10$~Gyr (green solid histogram), and $>10$~Gyr (red histogram). AD increases with stellar age.\label{fig:AD_whole_pop}}
\end{center}
\end{figure*}

To calculate the cumulative rotation curves and AD values as a function of stellar age for the whole primary sample, we first calculate both the rotation curve and AD for each individual analog as described in the previous section (see Figure~\ref{fig:analog_ad}). Since the rotation curves are radially binned, the cumulative rotation curve for the whole primary sample is constructed by taking the median rotation velocity for each age group along every radial bin. To compare the AD values across all of the analogs, we find the median of AD for each of the four stellar age bins for every analog. Thus, for our primary sample, we have 93 median AD values for each of the four age bins. We show the distribution of these median AD measurements along with the cumulative rotation curves across the primary sample of analogs in Figure~\ref{fig:AD_whole_pop}. We find there is a progression to higher AD as stellar age increases. This trend and the amplitude of median AD values are aligned with expectations from theory and with the observational study of AD as a function of stellar age in M31 \citepalias{Quirk2019}. The peak and the width of the distributions ($16^{\rm th} - 84^{\rm th}$ percentile) for each age bin are listed in Table~\ref{tab:whole_pop}. \citetalias{Quirk2019} find that stars that lag the gas in M31 tend to do so by $30~\pm10~\%$ \citepalias{Quirk2019}, while in the MW stars lagged behind the gas by $11~\pm8~\%$ \citep{Bovy_2009, Bovy_2012}. The AD in the MW is similar to that in other local galaxies \citep[e.g.][]{Ciardullo2004,Herrmann2009,Westfall2007, Westfall2011}. These observations probed stars that were younger than Group 4 in this analysis.

\begin{table}
\begin{center}
\label{tab:whole_pop}
\caption{Median AD Values for the Primary Sample}
\begin{tabular}{lc}
     \hline
     Stellar Age Group & Median AD (\kms) \\
     \hline
     Group 1: $< 1$~Gyr & $-6^{+5}_{-6}$ \\[2.5mm]
     Group 2: $1-5$~Gyr & $+10^{+8}_{-5}$ \\[2.5mm]
     Group 3: $5-10$~Gyr & $+56^{+16}_{-7}$ \\[2.5mm]
     Group 4: $>10$~Gyr & $+103^{+18}_{-33}$ \\
     \hline
\end{tabular}
\end{center}
\end{table}

\par
\begin{figure*}
\begin{center}
\includegraphics[scale=0.5]{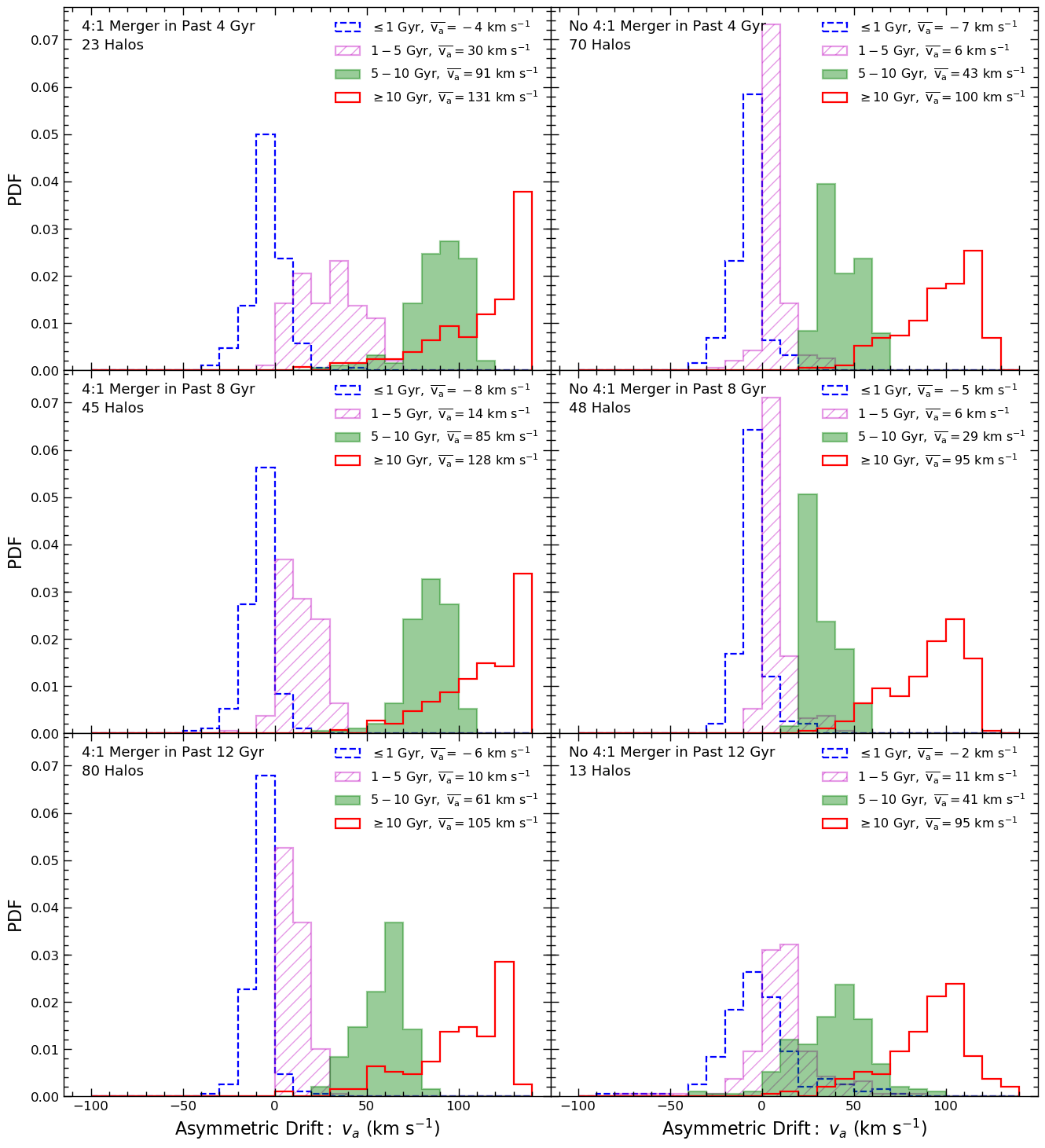}
\caption{Histograms of median AD values for the primary sample of analogs subdivided into two subgroups: those that have had a 4:1 merger (left panels) and those that have not (right panels) in a given time frame. In the top two panels, the time frame is the past 4~Gyr, in the middle panels it's the past 8~Gyr, and in the bottom panels it's the past 12~Gyr. In every panel, the blue histogram stars that are $< 1$~Gyr, the one with the purple hatch marks represents stars between $1-5$~Gyr, the green solid histogram represents stars between $5-10$~Gyr, and histogram denoted by the red line represents the stars $>10$~Gyr. The number of analogs in each subgroup is written in the top left of each panel. On average, the median AD is higher for the merger subgroups. \label{fig:merger_ad}}
\end{center}
\end{figure*}

 Since we aim to understand the possible correlation between AD and merger histories in this work, we use the properties included in the stellar assembly catalogues described in Section \ref{subsec:stellarassembly} to divide our primary sample into two subgroups: analogs that have experienced a 4:1 merger since a given time and those that have not experienced a merger more recent than the given time. We vary the time frame of interest from the past 4~Gyr, to the past 8~Gyr (approximately the median time of last 4:1 merger across the analogs), and to the past 12~Gyr. We create rotation curves (Figure~\ref{fig:merger_rcs}) and histograms of the median AD (Figure~\ref{fig:merger_ad}) for the subgroups for each of the specified time frames. The median AD values are listed in Table~\ref{tab:AD_merger}.
\par In the majority of cases for all time frames, AD is higher in the group that experienced the 4:1 merger than it is in the group that did not have a 4:1 merger more recent than the specified time frame (the ``no 4:1 merger group''). magentaThe exception is for Group 1 in the 4~Gyr time frame, where the magnitude of AD is slightly higher for the no 4:1 merger group and Group 2 in the 12~Gyr time frame, where AD is roughly equal for the two subgroups. For all time frames, the largest difference between the two subgroups is in Group 3. These are stars with ages $5-10$~Gyr. \citet{Brook_2004} show that the epoch in which a star is born can shape its kinematics. For example, they find stars born during a period of rapid hierarchical growth, roughly 8.5 to 10.5~Gyr ago, have the kinematical characteristics of thick disk stars, whereas those born in a quiescent epoch have characteristics that resemble the thin disk. Overall, the total difference in AD between the two subgroups decreases as the time since the last 4:1 merger increases, suggesting that some effects from ancient mergers settle over time.

\par In all cases, AD also increases with stellar age despite the time of last 4:1 merger. This is particularly interesting when looking at analogs that have not had a 4:1 merger in the past 12~Gyr. For these simulated galaxies, AD must be influenced by other dynamical heating sources, including internal sources of dynamical heating, such as GMCs, galactic bars, and spiral arms magenta or by other external events such as minor mergers and other tidal encounters. Another possible explanation is that stars were born in hotter and less rotationally-supported disks in the past or that the older stars were accreted. In the next section, we will discuss these implications further.

\begin{table}
\centering
\label{tab:AD_merger}
\caption{Median AD Values as a Function of Merger History}
\begin{tabular}{lccc}
     \hline
     Time Since Last & Stellar Age Group & Median AD  & Median AD \\
     4:1 Merger& & [\kms] & [\kms] \\
      & & (4:1 Merger) & (No 4:1 Merger) \\
     \hline
     4~Gyr & Group 1 & $-4^{+8}_{-8}$ & $-7^{+6}_{-8}$ \\[2.5mm]
     & Group 2 & $+30^{+18}_{-19}$ & $+6^{+5}_{-3}$\\[2.5mm]
    & Group 3 & $+91^{+13}_{-15}$ & $+43^{+13}_{-12}$\\[2.5mm]
     & Group 4 & $+131^{+14}_{-39}$ & $+100^{+16}_{-27}$\\
     \hline
      8~Gyr & Group 1 & $-8^{+6}_{-8}$ & $-5^{+6}_{-7}$ \\[2.5mm]
     & Group 2 & $+14^{+10}_{-11}$ & $+6^{+6}_{-4}$\\[2.5mm]
    & Group 3 & $+85^{+9}_{-13}$ & $+29^{+16}_{-5}$\\[2.5mm]
     & Group 4 & $+128^{+14}_{-33}$ & $+95^{+15}_{-28}$\\
     \hline
      12~Gyr & Group 1 & $-6^{+4}_{-6}$ & $-2^{+17}_{-16}$ \\[2.5mm]
     & Group 2 & $+10^{+7}_{-5}$ & $+11^{+12}_{-11}$\\[2.5mm]
    & Group 3 & $+61^{+9}_{-17}$ & $+41^{+15}_{-24}$\\[2.5mm]
     & Group 4 & $+105^{+21}_{-35}$ & $+95^{+15}_{26}$\\
     \hline
\end{tabular}
\end{table}

\section{Discussion}
\label{sec:Discussion}

\subsection{Comparisons to Measurements of Asymmetric Drift in Andromeda (Q19)}
\par
\begin{figure}
\begin{center}
\includegraphics[width=\columnwidth, scale=7.3]{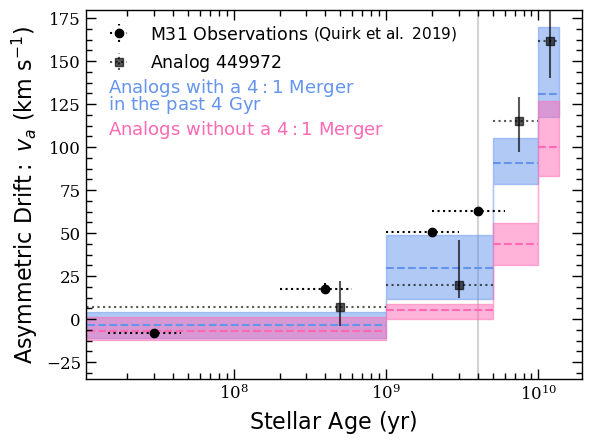}
\caption{AD as a function of stellar age for the primary sample of analogs and observations of M31's disk. The shaded regions represent the median AD of stellar Groups 1, 2, 3, and 4 from the primary sample of analogs. The dashed lines in each region shows the median AD value, and the shading shows the $1\sigma$ confidence levels. The blue represents analogs that have experienced a 4:1 merger within the past 4~Gyr, and the pink represents those that have not. The vertical line marks 4~Gyr. The black squares represent the median AD for the example analog 449972. This analog last had a 4:1 merger 4.12~Gyr ago. The observations from \citetalias{Quirk2019} are the black circles. The observations are most similar to the AD from the analogs that have experienced a 4:1 merger within the past 4~Gyr. \label{fig:comp_plot}}
\end{center}
\end{figure}

 We further examine the AD of analogs that have had a 4:1 merger in the past 4~Gyr and those that have not (top panels in Figures~\ref{fig:merger_ad} and \ref{fig:merger_rcs}), since recent studies suggest that M31 might have had a major merger in the past several billion years. Figure~\ref{fig:comp_plot} shows the comparison of AD in the simulated analogs to the AD measured in M31 as in \citetalias{Quirk2019}. The blue shaded regions represent AD for the analogs that have experienced a 4:1 merger in the past 4~Gyr, and the pink represents the AD for analogs that have not. The vertical line marks 4~Gyr. Stars to the right of this line existed during the merger event, while stars to the left might have been born during or after the merger. 

 \par The smallest difference in AD between the subgroups is in Group 1. These stars would have been born after the merger event. The largest difference is in Group 3. These stars would have formed $1-6$~Gyr before the merger. In all cases, AD is higher for the group of analogs that experienced a merger since a given time, showing that recent 4:1 mergers can perturb stars even in the inner 20~kpc of a galaxy.

 \par The black circular points in this figure show the observations of AD from \citepalias{Quirk2019}. Each point represents an average stellar age for a broad stellar evolution classification: massive main sequence (MS), intermediate mass asymptotic branch (AGB), older AGB, and less massive red giant branch (RGB). To emphasize that each point represents a range of stellar age, we have arbitrarily added dotted horizontal bars on each point to show +/- 50\% of the mean age. The star formation history of M31 is known more precisely than is represented here \citep{Lewis2015, Williams2017}, but we use broader age bins as to match the analysis in \citetalias{Quirk2019} and allow for easy comparison. The observational AD measurements from M31 are most similar to the blue shaded region, suggesting that the observed AD trends in M31 could be consistent with a 4:1 merger in the past 4~Gyr. We posit that internal heating could be an explanation for why the observed AD measurements are greater than the median values across the analog sample (see Section \ref{ssec:minmergers}). Furthermore, the observational M31 AD measurements are more similar to this analog subgroup than AD measured in analogs with more ancient mergers (the middle and bottom panels of Figure~\ref{fig:merger_ad}.) This is a potential piece of evidence that M31 did in fact have a recent major merger and suggests that the GSS could indeed have been formed by a major merger event that also led to prominent features that have been observed in M31's halo such as the various stellar streams, shelves, and tidal features mentioned in \cite{Hammer2018, DSouza2018}.  
 
For comparison, we also plot the median AD values for the example analog 449972 in black square points. The properties of this analog are shown in Table~\ref{tab:analog} and Figures~\ref{fig:analog_map} and \ref{fig:analog_ad}. This analog last had a 4:1 merger 4.12~Gyr ago, but its AD is much higher for the oldest stellar age groups than the overall primary sample. Since this analog has one of the highest $V_{\rm max}$ and $M_{\star}$ values in our sample, it may provide the best insight on the kinematics of the old stars in M31 relative to the younger stellar populations. 

\subsection{Assessing the Influence of Minor Mergers and Other Effects on AD}
\label{ssec:minmergers}
\par The AD observed in M31 naturally includes minor mergers, other tidal fly-bys, and sources of internal heating. The resolution of IllustrisTNG-100 does not allow us to probe how AD is affected by internal sources, like scattering via GMCs and spiral arms, as many internal heating sources rely on a thin disk. However, we can assess whether minor mergers play a significant role in the magnitude of AD and in the increasing AD as a function of stellar age using the primary sample.

\par  To check whether  minor mergers can also influence AD, we find five analogs in the primary sample with no 4:1 merger in the past 13.1~Gyr and analyze their dynamics based on the number of 10:1 mergers they have experienced. These five halos are split into a subgroup containing the halos that have experienced five to six 10:1 mergers and those that have only experienced zero or one 10:1 merger. AD increases with stellar age for all five halos. The AD for stellar age Groups 1, 2, and 3 is independent of the number of 10:1 mergers. The AD for the oldest stars (Group 4) is higher ($\rm \sim20~km\ s^{-1}$) in the analogs with the most 10:1 mergers but is still significantly below that of the AD for the primary sample, indicating that 4:1 mergers play a much larger role in influencing AD than 10:1 mergers. Furthermore, this suggests that minor mergers cannot explain why AD increases with stellar age (see also Section~\ref{sec:Results}, Figure~\ref{fig:merger_ad}) but rather that this pattern is likely a signature of any of the following processes: physical sources of internal heating, older stars being born with hotter birth kinematics, or from older stars born ex-situ. AD can be as high as $\rm 65~km\ s^{-1}$ for stellar age Group 4 even for analogs with zero or one 10:1 merger and no 4:1 mergers.Understanding the specifics of internal heating events versus artificial noise and their separate contributions to AD requires further work with a set of higher resolution simulations and is beyond the scope of this analysis.

\section{Summary and Conclusions}
\label{sec:Summary}
We have analyzed the AD of M31 mass analogs in the IllustrisTNG simulation to look for trends with stellar age. For the 93 analogs in our primary sample, we also examined merger histories to look for any correlations with AD. Our main conclusions are summarized below: 

\begin{description}
\item 1. We find that AD for M31 mass analogs in the IllustrisTNG simulation increases with stellar age, as in the observations in \citepalias{Quirk2019}. Our stellar age bins are defined as Group 1: $<1$ Gyr, Group 2: 1-5 Gyr, Group 3: 5-10 Gyr, and Group 4: $>10$ Gyr. Each stellar age bin is compared to the properties of neutral gas, defined as gas cells with neutral hydrogen fractions > 0.6.

\item 2. The M31 mass analogs are selected independently from their merger histories. Upon examining their merger trees, we find that major mergers do affect the value of AD. However, they are not the sole agent giving rise to observed trends in AD as we find that AD increases as a function of stellar age even without a major merger event. Thus, AD is also influenced by internal dynamical heating, minor mergers and birth kinematics. Understanding the specific internal processes by which AD is affected requires a suite of higher resolution zoom-in simulations such as Auriga or IllustrisTNG50.  

\item 3. We divide the primary sample into two subgroups based on the analogs that have experienced a 4:1  merger since a given time (i.e, in the last 4 Gyr, 8 Gyr, and 12 Gyr) and those that have not experienced a merger more recent than the given time to determine how strongly AD is affected by the most recent major merger. On average, analogs that have experienced a merger have higher AD (up to a $\rm 48~km\ s^{-1}$ difference) than those analogs that have not experienced a merger in the time frame of interest. The effect is greatest for stars in Group 3: those with age $5-10$~Gyr, regardless of time since the last 4:1 merger.

\item 4. The AD measurements observed in the disk of M31 from \citetalias{Quirk2019} are similar to the subgroup of the primary sample that experienced a 4:1 merger within the past 4~Gyr. This is a piece of evidence that M31 may have experienced a major merger event in the past several billion years.

\end{description}

\section*{Acknowledgements}
ACNQ is supported by National Science Foundation through the Graduate Research Fellowship Program funded by Grant Award No. DGE-1842400. EP was supported by the National Science Foundation through the Graduate Research Fellowship Program funded by Grant Award No. DGE-1746060 and is currently supported by the Miller Institute for Basic Research, University of California Berkeley. The authors would like to thank Julianne Dalcanton for bringing them together during the PHAT team meeting at Ringberg Castle in 2018 for without it, this project may not have emerged. The authors would also like to thank Greg Snyder for the insightful and supportive conversations in the initial stages of this project and Kathryn Johnston and Puragra Guhathakurta for the helpful discussions and comments. Thank you to Vicente Rodriguez-Gomez for analyzing the assembly history of IllustrisTNG galaxies and for making the stellar assembly catalogs available for this project. The authors would like to thank the anonymous referees for their feedback which has greatly improved the quality of this project. The authors acknowledge the significance and importance of the summit of Mauna Kea to the indigenous Hawaiian community and appreciate the opportunity to use data collected from this sight for this analysis.\\

\noindent {\it Software:} This research made use of \texttt{astropy} \citep{astropy2013, astropy2018}, \texttt{Illustris Python} \citep{nelson15}, \texttt{matplotlib} \citep{Hunter:2007}, \texttt{numpy} \citep{numpy}, and \texttt{scipy} \citep{scipy}. \\

\noindent {\it Data Availability Statement:} The data used in this article is from the IllustrisTNG project and is available upon request at https://www.tng-project.org/data/. 

\bibliographystyle{mnras}
\bibliography{Mendeley} 

\newpage
\appendix
\section{Particle Density Plots for Example Analog 449972}
\label{sec:appendixA}

\begin{figure*}
    \centering
    \includegraphics[scale=0.65]{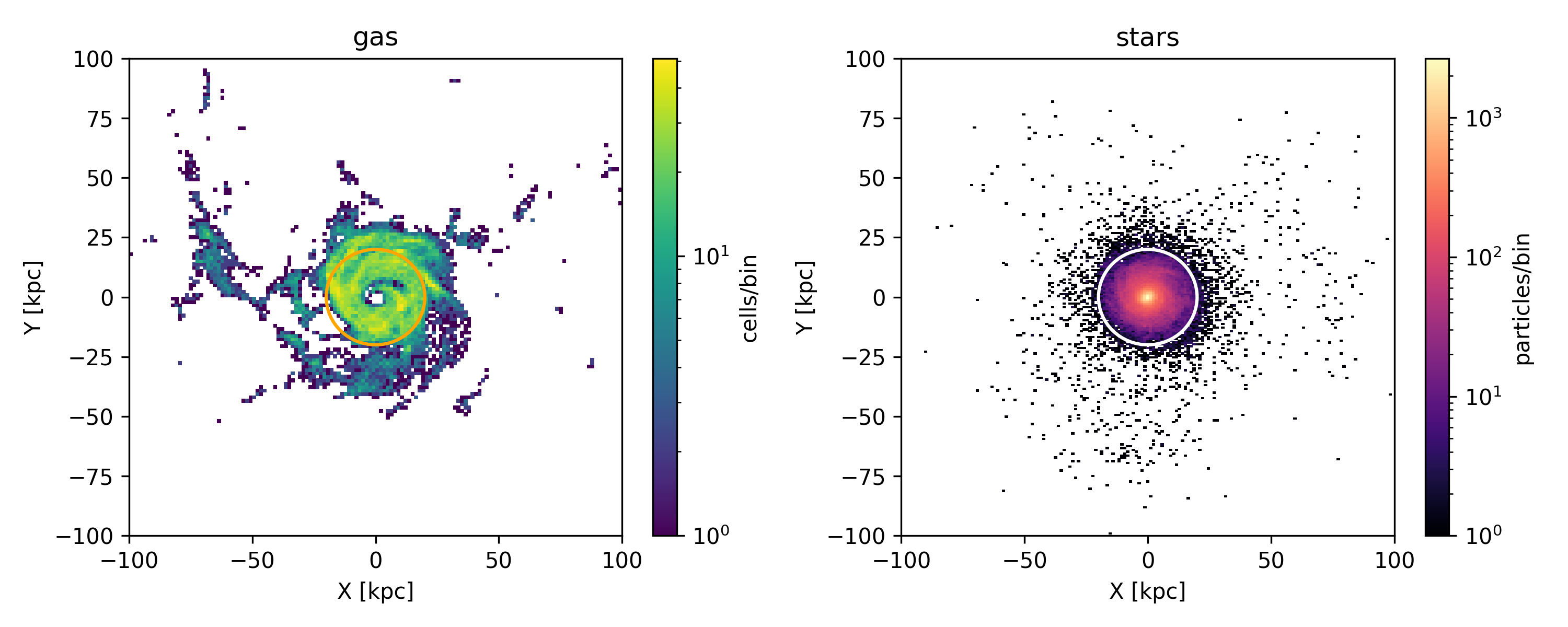}
    \caption{Density maps for one M31 analog (ID 449772). The data has been rotated to the face-on reference frame as described in Section \ref{subsec:rotcoord} and only includes particles within 10 kpc of the disk plane. The left panel shows the gas density only for the cells where $n_H>0.6$ and the right panel shows the stellar density. In the left and right panels, the orange and white circles, respectively, denote a radius of 20 kpc. }
    \label{fig:3Dmap}
\end{figure*}

Figure~\ref{fig:3Dmap} shows a more detailed view of analog 449972's geometry. The left panel shows the gas cell density in the inner 100 kpc of the simulated galaxy. The right panel shows the stellar particle density across the same extent. The orange and gray circles illustrates the 20 kpc region encompassing the data in the left and right panels, respectively. All rotation curves are calculated using particle/cell data within this 20 kpc region relative to the COM. For this M31 analog, 56\% of gas cells with $n_H > 0.6$ and 97\% of stellar particles are encompassed by the 20 kpc radius, so the rotation curves used to calculate AD are representative of the most prominent inner regions of the M31 analogs with little contamination from non-spherical substructures.

\section{Primary Sample Rotation Curves}\label{sec:appendixB}
\begin{figure*}
\begin{center}
\includegraphics[scale=0.5]{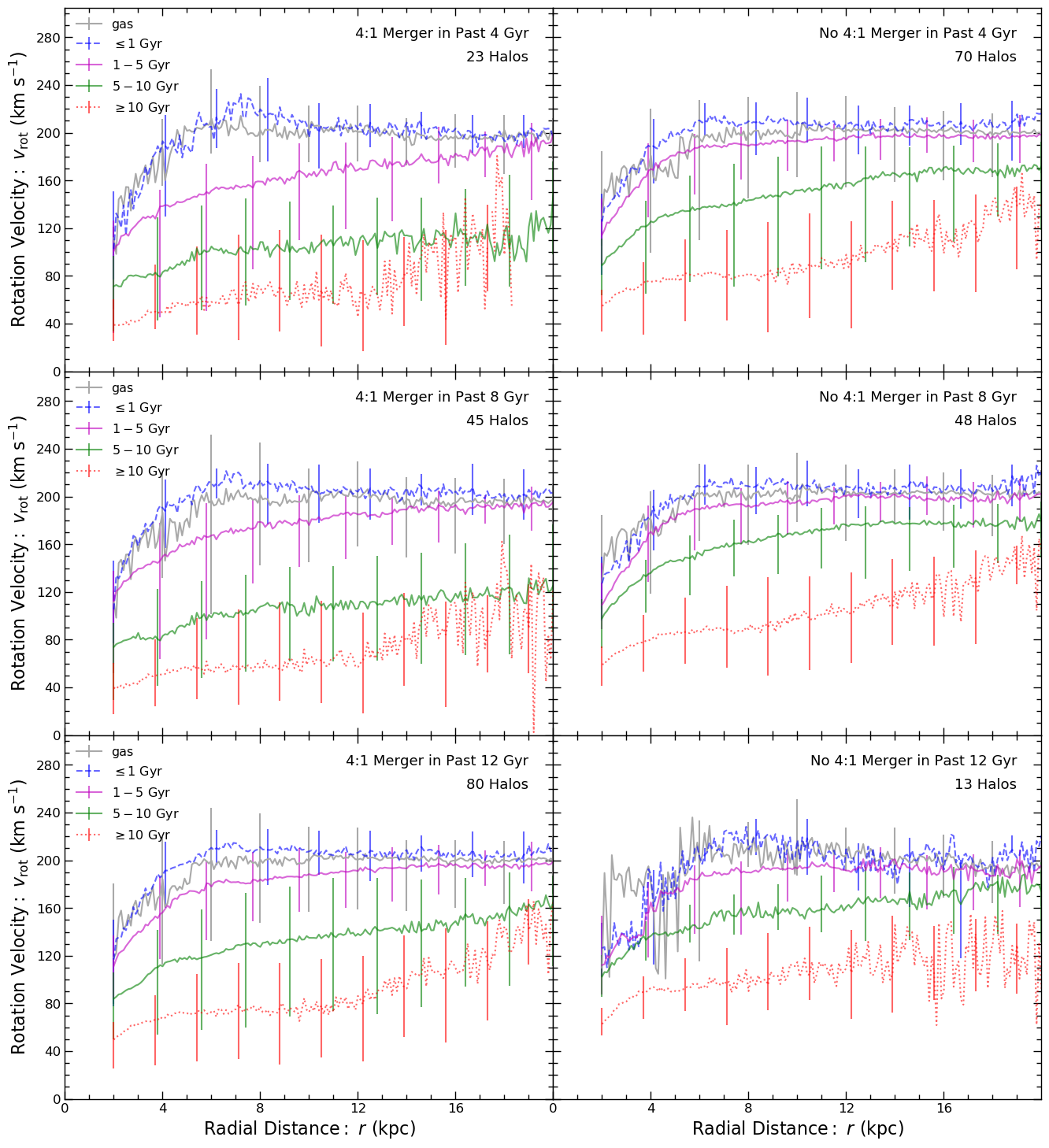}
\caption{Rotation curves for the primary sample of analogs subdivided into two groups: those that have had a recent 4:1 merger (left panels) and those that have not (right panels). In the top two panels, we define 4:1 merger time frame as having occurred in the last 4~Gyr. In the middle panels, the time frame is 8~Gyr, and in the bottom panels it is 12~Gyr. In every panel, the grey line shows the rotation curve of the neutral gas, the blue dashed line represents stars that are $< 1$~Gyr, the purple represents stars between $1-5$~Gyr, the green represents stars between $5-10$~Gyr, and the red dotted line represents the stars $>10$~Gyr. The vertical bars show the width of distribution of median rotation values ($16^{\rm th} - 84^{\rm th}$ percentile) across all of the analogs at a given radial bin. The oscillations in the individual rotation curves shows the radial variations in rotation velocity.   \label{fig:merger_rcs}}
\end{center}
\end{figure*}

In this section, we show the rotation curves from the merger and no merger subgroups (Figure~\ref{fig:merger_rcs}). These rotation curves are used to make the histograms of AD shown in Figure~\ref{fig:merger_ad}. The vertical gap between the stellar particle rotation curves and the gas data rotation curve is a visual representation of AD. 

\label{lastpage}
\end{document}